\documentclass[useAMS,usenatbib]{mn2e}
\usepackage{graphicx} 
\newcommand{\be}{\begin{equation}}

\newcommand{\ee}{\end{equation}}
\newcommand{\bea}{\begin{eqnarray}}
\newcommand{\eea}{\end{eqnarray}}
\newcommand{\bef}{\begin{figure}}
\newcommand{\eef}{\end{figure}}
\newcommand{\bce}{\begin{center}}
\newcommand{\ece}{\end{center}}
\def\lsim{\mathrel{\rlap{\lower4pt\hbox{\hskip1pt$\sim$}}
    \raise1pt\hbox{$<$}}}         
\def\gsim{\mathrel{\rlap{\lower4pt\hbox{\hskip1pt$\sim$}}
    \raise1pt\hbox{$>$}}}         

\title[Measured neutron star masses and lattice QCD]
{A link between measured neutron star masses and lattice QCD data}

\author[I. Bombaci and D. Logoteta]{Ignazio Bombaci$^{1}$\thanks{E-mail:bombaci@df.unipi.it} 
and Domenico Logoteta$^{2}$ \\
$^{1}$Dipartimento di Fisica `Enrico Fermi', Universit\'a di Pisa,
and INFN Sezione di Pisa,  Largo Bruno Pontecorvo 3, I-56127 Pisa, Italy\\
$^{2}$Centro de F\'{\i}sica Computacional, Department of Physics, University of Coimbra, 
3004-516 Coimbra, Portugal}
\begin{document}

\date{Accepted 2013 May 16. Received 2013 May 14; in original form 2013 January 29}

\pagerange{\pageref{firstpage}--\pageref{lastpage}} \pubyear{2013}

\maketitle

\label{firstpage}

\begin{abstract}
We study the hadron--quark phase transition in neutron star matter and the structural 
properties of hybrid stars using an equation of state (EOS) for the quark phase derived with 
the field correlator method (FCM). 
We make use of the measured neutron star masses, and particularly the mass of PSR~J1614$\textendash$2230,  
to constrain the values of the gluon condensate $G_2$ which is one of the EOS parameter 
within the FCM. 
We find that the values of $G_2$ extracted from the mass measurement of PSR~J1614$\textendash$2230 
are fully consistent with the values of the same quantity derived, within the FCM, 
from recent lattice quantum chromodynamics (QCD) calculations of the deconfinement transition temperature at 
zero baryon chemical potential. 
The FCM thus provides a powerful tool to link numerical calculations of QCD on 
a space--time lattice with neutron stars physics. 
\end{abstract}


\section{Introduction}
Neutron stars, the compact remnants of supernova explosions, are unique natural 
laboratories to explore the phase diagram of quantum chromodynamics (QCD) 
in the low-temperature $T$ and high-baryon-chemical-potential $\mu_b$   
region \citep{web05,alf08}. 
In this regime non-perturbative aspects of QCD are expected to play a crucial role,   
and a transition to a phase with deconfined quarks and gluons is expected to occur and 
to influence a number of interesting astrophysical 
phenomena \citep{pg10,sot11,r1,r3,r6}. 

Recent high-precision numerical calculations of QCD on a space--time lattice at 
$\mu_b = 0$ (i.e. zero baryon density) have shown that at high temperature 
and for physical values of the quark masses, the transition to quark gluon plasma is a 
crossover \citep{aoki06} rather than a real phase transition.  

Unfortunately, present lattice QCD calculations at finite baryon chemical potential 
are unrealizable by all presently known lattice methods (see e.g. \citet{mp_lomb08}). 
Thus, to explore the QCD phase diagram at low T and high $\mu_b$, it is necessary 
to invoke some approximations in QCD or to apply some QCD effective model.   

Along these lines, different models of the equation of state (EOS) of quark matter,  
as the bag model \citep{mit-eos} or the Nambu--Jona-Lasinio (NJL) model \citep{njl,bub05},   
have been intensively used by many authors to calculate the structure of 
strange stars \citep{witt84,afo86,hzs86}, or the structure of the 
so called hybrid stars, {\it i.e.} neutron stars with a quark matter core 
(see e.g \citet{klahn07,lenzi12}).   
These EOS models are expected to be reasonable at very large density,  but 
they crumbles in the density region where quarks clusterize to form hadrons, i.e. 
in the region where the deconfinement phase transition takes place. 
In addition, the bag model and the NJL model, as other QCD effective models,   
can not make predictions in the high $T$ and zero $\mu_b$ region, and thus 
cannot be tested using present lattice QCD calculations. 

A promising approach to describe the EOS of the quark gluon phase is the so called 
Polyakov loop extended NJL model 
\citep{meisinger96,fukushima04,ratti06,blaschke08,contrera08,blaschke10,dexh10},  
which combines the two main nonperturbative aspects of low energy QCD: 
confinement and spontaneous chiral symmetry breaking.

Recently the deconfinement phase transition has been described using an EOS of quark 
gluon plasma derived within the field correlator method (FCM) \citep{st,digiac02}  
extended to finite baryon chemical potential \citep{ST07,sim1,sim3,sim5}. 
The FCM is a nonperturbative approach to QCD which includes from first principles the 
dynamics of confinement. The model is parametrized in terms of the gluon condensate $G_2$ 
and the large distance static $Q\bar{Q}$ potential $V_1$.  These two parameters 
control the EOS of the deconfined phase at fixed quark masses and temperature. 
The main constructive characteristic of the FCM is the possibility of describing the whole 
QCD phase diagram  as it can span from high temperature and low baryon chemical potential,  
to low $T$ and high $\mu_b$ limit. 

A very interesting feature of the FCM is that the value of the gluon condensate can 
be obtained  from lattice QCD calculations of the deconfinement transition temperature 
$T_c$, at zero baryon chemical potential. Thus, we have an efficacious tool 
to directly link lattice QCD simulations and neutron star physics.    

To explore this link is the main purpose of this work.   
In particular, we will investigate the possibility for the occurrence of the quark 
deconfinement transition in neutron stars and the possibility of having stable hybrid star 
configurations using the FCM for the quark phase EOS and a relativistic mean field 
model \citep{gm} for the EOS of the  hadronic phase. 
  
\section{EOS of the Quark Phase}
\label{2}
The quark matter EOS we used in this work is based on the FCM \citep{st,digiac02}.   
Recently, this method has been extended to the case of non-zero baryon 
density \citep{ST07,sim1,sim3,sim5} making possible its application to neutron star matter. \\
The main advantage of the FCM is a natural explanation and treatment of the dynamics of 
confinement in terms of colour electric $D^{E}(x)$, $D_{1}^{E}(x)$ and colour magnetic 
$D^{H}(x)$, $D_{1}^{H}(x)$ Gaussian correlators \citep{digiac02}.   \\
$D^{E}$ contributes to the standard string tension $\sigma^{E}$ through \citep{ST07}: 
\be
\label{f1}
              \sigma^{E}=\frac{1}{2}\int D^{E}(x) \ d^{2}x .
\ee
The string tension $\sigma^{E}$ vanishes as $D^{E}$ goes to zero at $T \geqslant T_c$,   
and this leads to deconfinement. The correlators have been calculated on the 
lattice \citep{pisa} and also analytically \citep{sim98}. 

Within the FCM the quark pressure $P_q$, for a single flavour, reads 
\citep{ST07}   
\be\label{pquark}
P_q/T^4 = \frac{1}{\pi^2}[\phi_\nu (\frac{\mu_q - V_1/2}{T}) +
\phi_\nu (-\frac{\mu_q + V_1/2} {T})]
\ee
where 
\be
\label{eq:phi}
\phi_\nu (a) = \int_0^\infty du \frac{u^4}{\sqrt{u^2+\nu^2}} \frac{1}{(\exp{[ \sqrt{u^2 +
\nu^2} - a]} + 1)} \, , 
\ee
$\nu=m_q/T$ and  $V_1$ is the large distance static $Q\bar Q$ potential:
\be
\label{v1}
V_1(T) = \int_0^{1/T} d\tau(1-\tau T) \int_0^\infty d\chi \chi D_1^E(\sqrt{\chi^2 + \tau^2})\, .
\ee
The non-perturbative contribution to  $D_1^E(x)$ is parametrized as \citep{digiac02}
\be
\label{d1nonpt}
D_1^E(x)=D_1^E(0) \exp(-|x|/\lambda)
\ee
where $\lambda$ is the vacuum correlation length.  
Following \citet{ST07}, we use the value  $\lambda = 0.34~\rm{fm}$ which has been 
determined in lattice QCD calculations \citep{d'elia97}. 

In this formalism, $V_{1}$ in equation(\ref{v1}) is independent of the chemical potential 
(and so on the density). This feature is partially supported by lattice simulations 
at small chemical potential \citep{ST07,latmuf}.  
In thi work, the value of $V_{1}$ at $T=0$ has been considered as a 
model parameter.   

The gluon contribution to the pressure is \citep{sim5}
\be
\label{pglue}
P_g/T^4 = \frac{8}{3 \pi^2} \int_0^\infty  d\chi \chi^3
\frac{1}{\exp{(\chi + \frac{9 V_1}{8T} )} - 1}   \, .
\ee

In summary the total pressure of the quark phase is given by 
\be
\label{pq}
      P_{qg} = P_{g} + \sum_{u,d,s} P_{q} - \frac{9}{64} G_2 \, .
\ee
The last term in equation(\ref{pq}) represents the vacuum energy difference between 
the quark and hadronic phases, in the case of three flavour ({\it u, d, s}) quark 
matter \citep{ST07}, and $G_2$ is the gluon condensate. 
The latter quantity has been determined, with large uncertainty, using QCD sum rules \citep{QCD-sum} 
$G_2 = (0.012 \pm 0.006)~{\rm GeV}^4$. 
In this work, the value of $G_2$ has been considered as a model parameter.   
We used the following values of the current-quark masses: $m_u = m_d = 5$~MeV and $m_s = 150$~MeV. 
In summary, the quark matter EOS has two parameters: $G_2$ and $V_1 = V_1(T=0)$.

\section{Neutron star structure}
\label{NSstr}

In this section, we show the results of our calculations of hybrid stars structure.  
To this purpose, we integrate the well-known Tolman, Oppenheimer and Volkov
relativistic hydrostatic equilibrium equations to get various stellar properties for a 
fixed EOS.  
For the hadronic phase, we consider $\beta$-stable nuclear matter, 
and we make use of a non-linear relativistic mean field model in the parametrization 
GM1 given by \citet{gm}.  
The GM1 model can be considered a representative realistic nuclear EOS in the sense that it 
fits the empirical saturation properties of nuclear matter, does not violate causality at high 
density and is compatible with the present measured neutron star masses.  
All the results presented in the following have been obtained using the Gibbs 
construction \citep{gle92} to model the hadron--quark phase transition.

\begin{figure}
\centering
\vskip 0.5cm
\includegraphics[width=1.0\columnwidth]{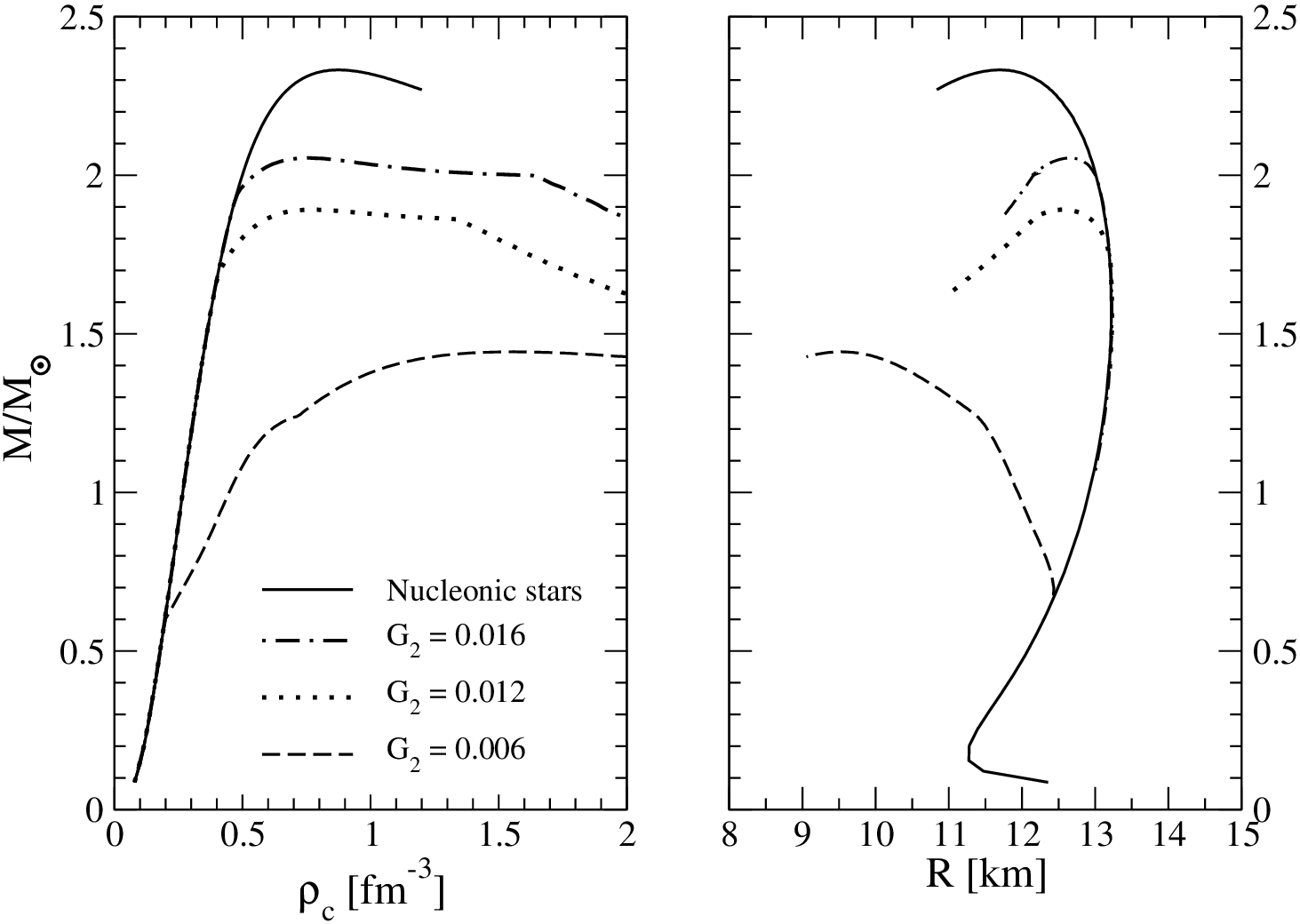}
\caption{Stellar gravitational mass $M$ versus central baryon number density 
$\rho_c$ (left-hand panel) and versus stellar radius $R$ (right-hand panel) for hybrid stars for 
several values of the gluon condensate $G_{2}$ (reported in GeV$^4$ units) 
and for $V_{1} = 0.01 \, {\rm GeV}$.   
The continuous line in both panels refers to the pure nucleonic stars, i.e. compact 
stars with no quark matter content.}
\label{Mrho+MR_N}
\end{figure}

In Fig. \ref{Mrho+MR_N} we report the stellar gravitational mass $M$ 
(in unit of the solar mass $M_\odot = 1.99 \times 10^{33} \rm{g}$) 
versus the central baryon number  density $\rho_{c}$ (left-hand panel) and the 
mass versus radius $R$ (right-hand panel) in the case of pure nucleonic stars (continuous line)  
and of hybrid stars for different $G_{2}$ and taking $V_{1} = 0.01 \, {\rm GeV}$.  
We obtain stable hybrid star configurations for all the considered values of the gluon 
condensate, with maximum masses ranging from  
$M_{max} = 1.44 \, M_\odot$ (case with $G_{2} = 0.006 \, {\rm GeV}^4$) to  
$M_{max} = 2.05 \, M_\odot$ ($G_{2} = 0.0016 \, {\rm GeV}^4$).   
Note that the hybrid star branch of the stellar equilibrium configurations shrinks as $G_{2}$ 
is increased. 
This behaviour is different with respect to the one found by \citet{Baldo},  
where the stability window of hybrid star configurations 
was restricted within the range $0.006 \, {\rm GeV}^4 < G_{2} < 0.007 \, {\rm GeV}^4$.\\   
The properties of the maximum mass configuration for hybrid star sequences varying  
$G_{2}$ are summarized in table 1.   

\begin{table}
\caption{Properties of the maximum mass configuration for hybrid stars as a function of the 
gluon condensate $G_{2}$.} 
  \label{table}
 \begin{tabular}{@{}cccc}
  \hline
 $G_2$~(GeV$^{4}$) ~~~~~~ & $M_{max}(M_\odot$) ~~~~ & $\rho_c^{Hyb}$~(fm$^{-3}$)~~~ & $R$~(km) \\
  \hline
   0.006 &    1.44  &    1.55 &     9.54 \\
   0.012 &    1.89  &    0.77 &    12.55 \\
   0.016 &    2.05  &    0.75 &    12.66 \\                
  \hline
 \end{tabular}

 \medskip
The results are relative to the case $V_{1} = 0.01 \, {\rm GeV}$. 
The maximum mass configuration for the pure nucleonic star sequence is:   
$M_{max} = 2.33\, M_\odot$, $\rho_{c}^{NS} = 0.87\, \rm{fm}^{-3}$ and $R = 11.70\, {\rm km}$.
\end{table}

\begin{figure}
\centering
\vskip 0.7cm
\includegraphics[width=1.0\columnwidth]{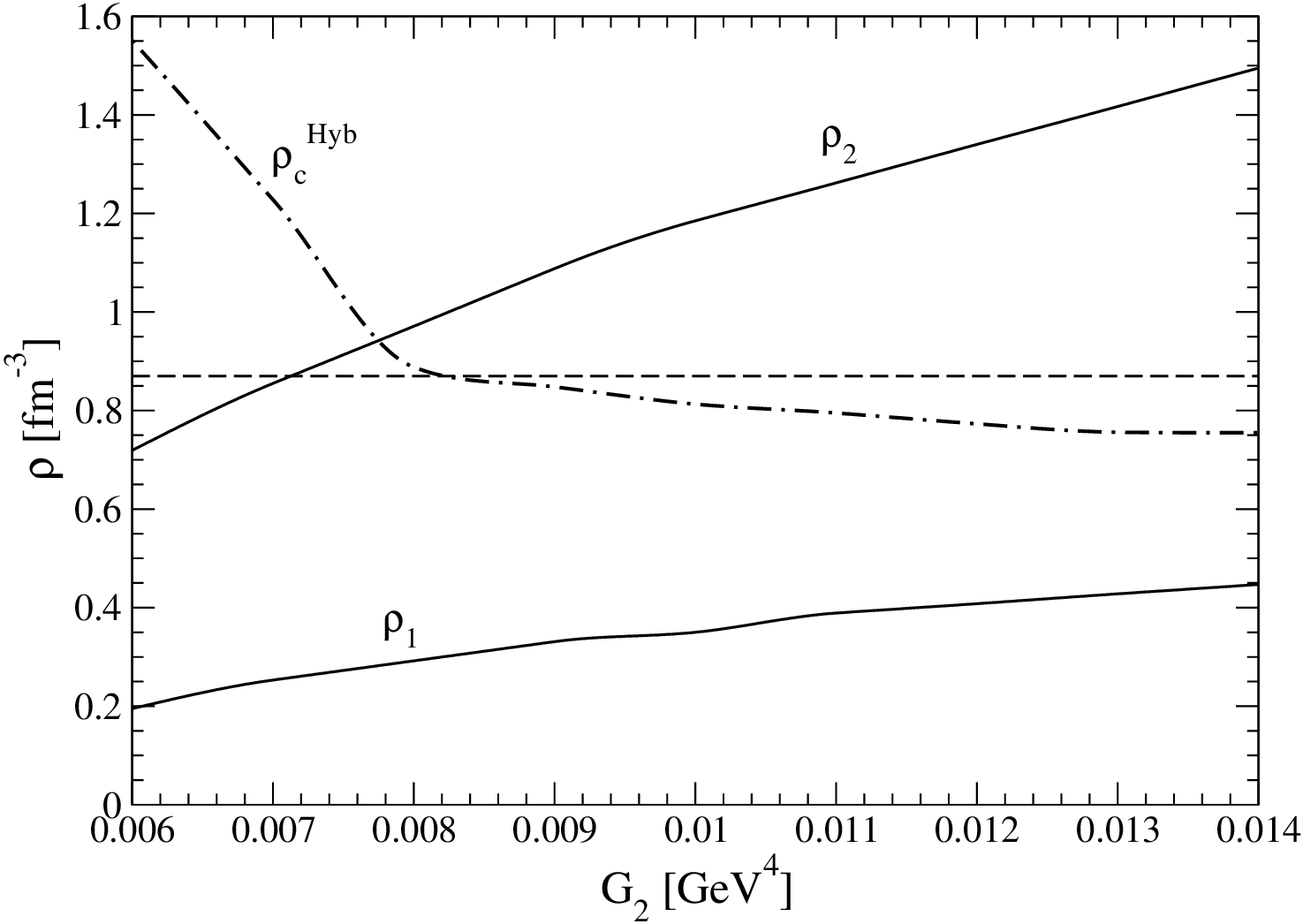}
\caption{Quark--hadron phase transition boundaries in $\beta$-stable nuclear matter 
as a function of the gluon condensate $G_2$ and for $V_{1} = 0.01 \, {\rm GeV}$. 
The onset of quark--hadron mixed phase occurs at the baryon number density $\rho_1$, 
and the pure quark phase begins at $\rho_2$.  Also shown is the central baryon number 
density $\rho_c^{Hyb}$ of the maximum mass hybrid star. The horizontal dashed line represents 
the value $\rho_c^{NS}$ of the central baryon number density of the maximum mass pure nucleonic star.}
\label{rho_g2_N}
\end{figure} 
In Fig. \ref{rho_g2_N}, we plot the quark--hadron phase transition boundaries in 
$\beta$-stable nuclear matter as a function of  $G_2$ and with $V_{1} = 0.01 \, {\rm GeV}$.     
The onset of the deconfinement transition (i.e. the onset of the 
quark--hadron mixed phase) occurs at the baryon number density $\rho_1$, 
and the pure quark phase begins at $\rho_2$.    
Also, shown is the central baryon number density $\rho_c^{Hyb}$ of the 
maximum mass hybrid star (dot--dashed line). 
Stable hybrid star configurations have central densities $\rho_c$ located in the region 
of the  $\rho$--$G_2$ plane between the lower continuous line and the dot--dashed line, 
i.e. $\rho_1 < \rho_c  \leqslant \rho_c^{Hyb}$.   
Note that $\rho_c^{Hyb} > \rho_2$ when the gluon condensate is in the range 
$0.006  \, \rm{GeV}^4 < G_{2}  \leqslant 0.0077  \, \rm{GeV}^4$.  
For these $G_{2}$ values all hybrid stars with a central density in the range 
$\rho_2 < \rho_c \leqslant  \rho_c^{Hyb}$  possess a pure quark matter core.  
Finally the horizontal dashed line represents the value $\rho_c^{NS}$ of the central baryon 
number density of the maximum mass pure nucleonic star.   

\begin{figure}
\centering
\vskip 0.7cm
\includegraphics[width=1.0\columnwidth]{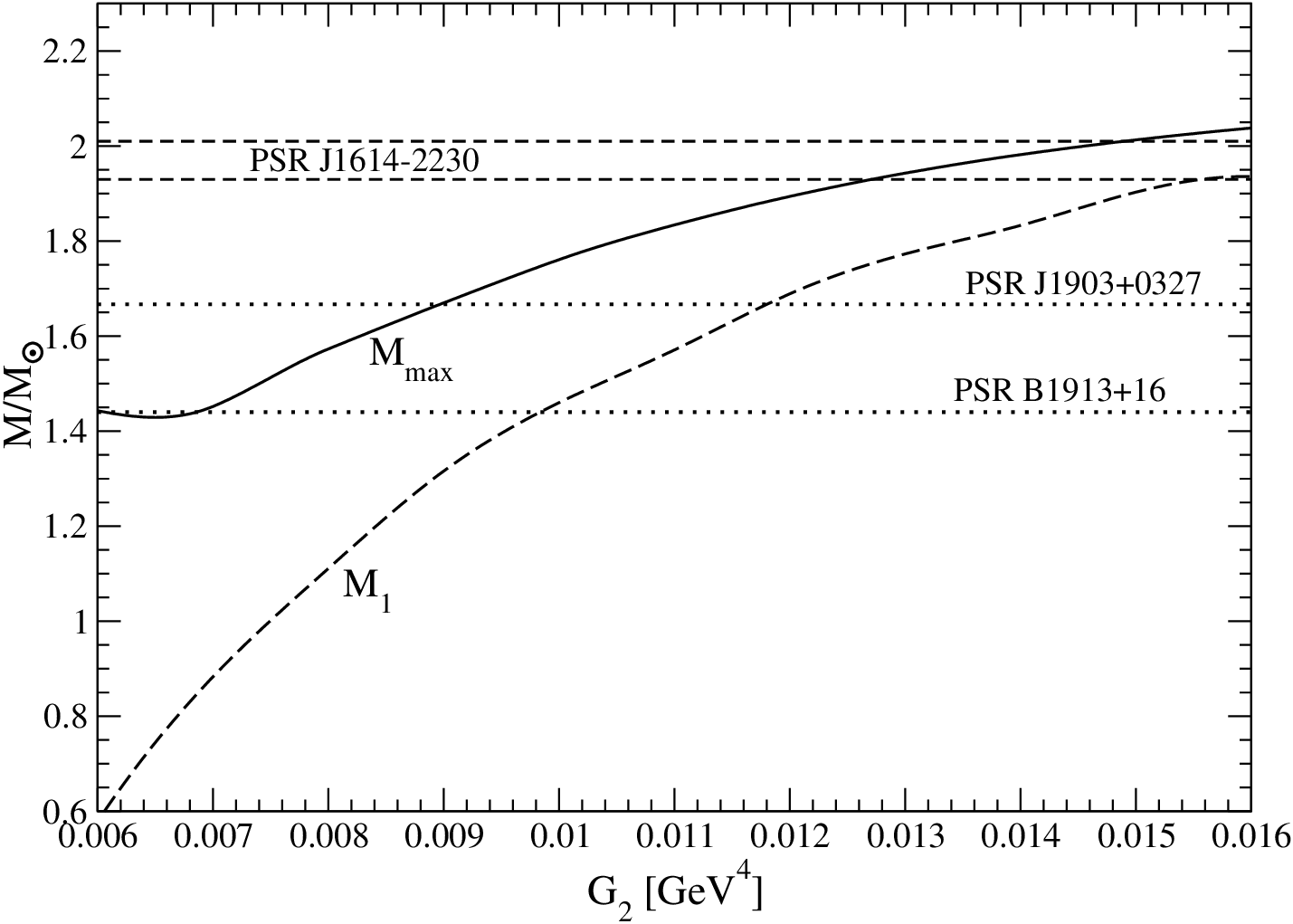}
\caption{Gravitational maximum mass for hybrid stars (continuous line) and gravitational 
mass $M_1$ (dashed line) of the star with central baryon number density $\rho_1$ corresponding 
to the onset of mixed quark--hadron phase as a function of the gluon condensate $G_{2}$ 
and for $V_{1} = 0.01 \, {\rm GeV}$.} 
\label{Mmax_g2_N}
\end{figure}

In Fig. \ref{Mmax_g2_N} we draw the maximum mass $M_{max}$ for hybrid stars 
(continuous line) and the mass $M_1 = M(\rho_1)$ (dashed line) of the star with central 
baryon number density $\rho_1$ corresponding to the onset of the mixed phase.    
These two quantities are plotted as a function of the gluon condensate $G_{2}$ 
and taking $V_{1} = 0.01 \, {\rm GeV}$. 
Stable hybrid star configurations correspond to the region of the $M$--$G_2$ plane 
between the continuous and the dashed line.    
Stellar configurations in the region below the dashed line $M_1$ do not possess any 
deconfined quark matter in their centre (pure nucleonic stars).    

To compare our results with measured neutron star masses, we report in 
the same Fig. \ref{Mmax_g2_N}, the values of the masses of the following pulsars: 
PSR~B1913+16  with $M = 1.4398 \pm 0.0002 \, M_\odot$ \citep{HT75,wnt10},        
PSR~J1903+0327 with $M = 1.667 \pm 0.021 \, M_\odot$ \citep{frei11} and  
PSR~J1614--2230  with $M = 1.97 \pm 0.04 \, M_\odot$ \citep{demo10}.  \\  

The mass of PSR~J1614--2230 gives the strongest constraint on the possible value 
of the gluon condensate.  In fact, using the lower bound of the measured mass of 
PSR~J1614--2230, we get  $G_{2} \geq  0.0129~{\rm GeV}^4$.
Thus, for values of the gluon condensate in the range  
$0.0129~{\rm GeV}^4  \leq G_{2} \leq G_{2}^{*} \simeq 0.018~{\rm GeV}^4$,    
PSR~J1614--2230 is a hybrid star,  whereas PSR~B1913+16 and PSR~J1903+0327 
are pure nucleonic stars. In the above specified range for the gluon condensate, 
$G_{2}^{*}$ is defined by the condition $M_1(G_{2}^{*}) = 2.01~M_\odot$, 
the upper bound of the measured mass of PSR~J1614-2230. 
Thus for $G_{2} > G_{2}^{*}$, PSR~J1614--2230 is a pure nucleonic star.  

To explore the influence of the large distance static $Q\bar{Q}$ potential 
on the stellar properties, we have considered a quark phase EOS with $V_1 = 0.10 \,{\rm GeV}$.    
Once again we get stable hybrid star configurations for all the considered values of $G_2$, 
with maximum masses ranging from  
$M_{max} = 2.00 \, M_\odot$ (case with $G_{2} = 0.006 \, {\rm GeV}^4$) to  
$M_{max} = 2.25 \, M_\odot$ ($G_{2} = 0.0016 \, {\rm GeV}^4$).  
Thus an increase of the value of $V_1$ reduces the extension of the hybrid star branch,   
shifts it to larger densities and produces hybrid stars with a larger maximum mass.  
In this case we found that the calculated $M_{max}$ is compatible with  
the lower bound of the measured mass of PSR~J1614--2230 for all the considered 
values of the gluon condensate (i.e.  $G_2  \geq 0.006 \, {\rm GeV}^4$). 

We also considered stellar models with $V_1 = 0.50 \, {\rm GeV}$ and 
$V_1 = 0.85 \, {\rm GeV}$. In these two cases no phase transition occurs in neutron 
stars (i.e. $\rho_1 > \rho_c^{NS}$ the central density of the maximum mass pure nucleonic star), 
thus in this case PSR~J1614--2230 would be a pure nucleonic star.

\section{Lattice QCD calculations and measured neutron star masses}
\label{Tc}   
\begin{figure}
\centering
\includegraphics[width=0.95\columnwidth]{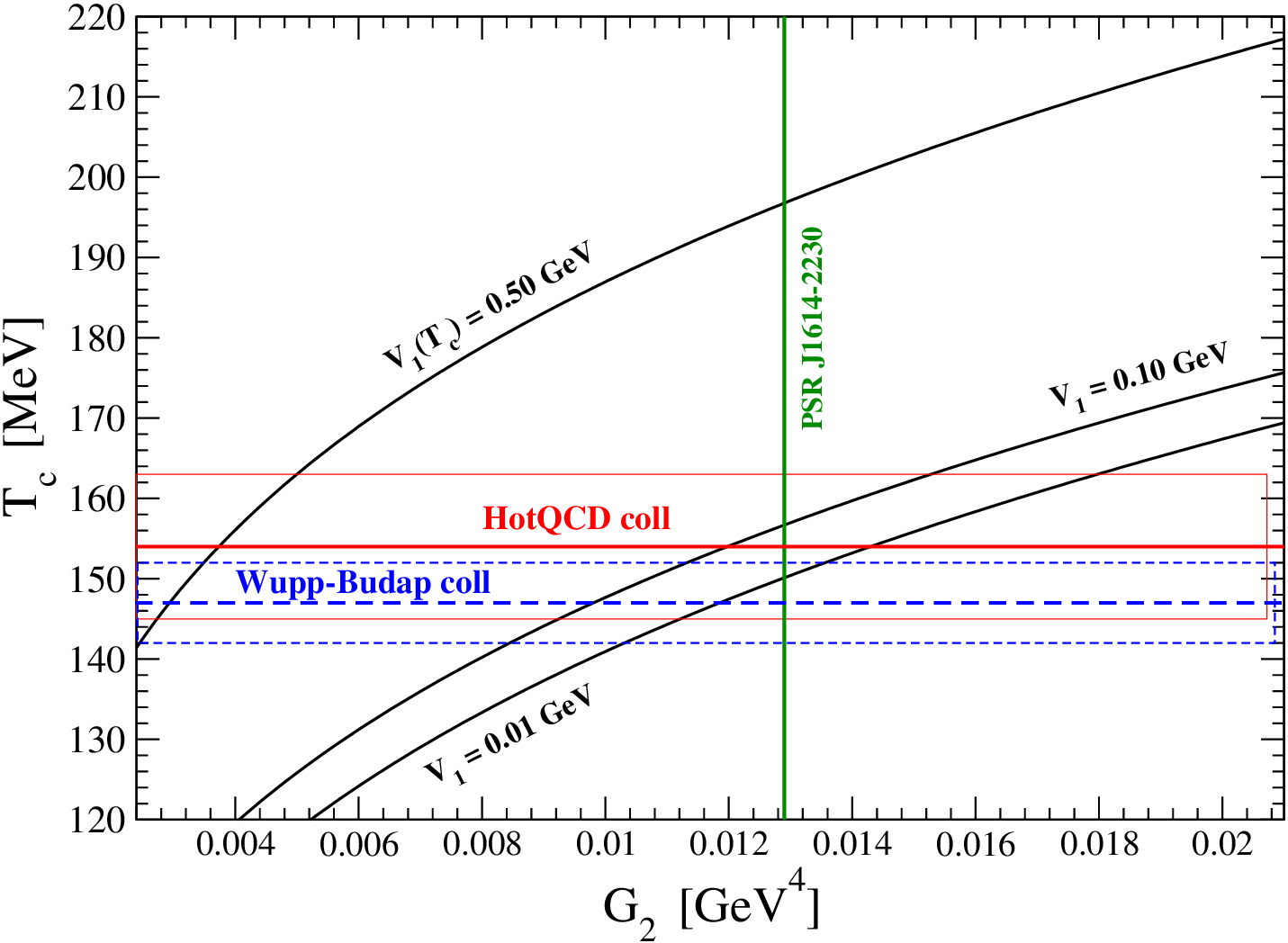}
\caption{(Deconfinement transition temperature $T_c$ at $\mu_b = 0$.   
The curve labeled with $V_1(T_c) = 0.5 \,{\rm GeV}$ reproduces the FCM results of 
\citet{ST07} for a fixed value $V_1(T_c) = 0.5 \,{\rm GeV}$ of the large distance 
static $Q\bar{Q}$ potential.  
The curve labelled with $V_1= 0.01\,{\rm GeV}$ ($V_1= 0.10\,{\rm GeV}$)  
corresponds to the transition temperature at $\mu_b = 0$ obtained   
solving numerically equations~(\ref{Tc0}) and  (\ref{v1_T}) for the case 
$V_1(0) = 0.01\,{\rm GeV}$ [$V_1(0) = 0.10\,{\rm GeV}$].      
The horizontal heavy and thin lines represent, respectively, the central 
value and the error estimate of lattice QCD calculations. 
In particular, the red continuous lines refer to the calculations~\citep{baz12} 
of the Hot QCD collaboration  $T_c = (154 \pm 9)\, {\rm MeV}$;  
the blue short dashed lines refer to the calculations~\citep{bors10}  
of the Wuppertal--Budapest collaboration  $T_c =  (147 \pm 5)\, {\rm MeV}$. 
Finally, the vertical green line represents the lower limit for $G_2$ which is compatible 
with the lower bound of the measured mass of PSR~J1614--2230 for the case 
$V_1(0) = 0.01\,{\rm GeV}$.  
}   
\label{Tc_G2}
\end{figure}

Within the FCM, the deconfinement transition temperature $T_c$ at $\mu_b = 0$ 
reads \citep{ST07} 
\be
\label{Tc0}
T_c = \frac{a_0}{2} G_2^{1/4} \Bigg( 1 + \sqrt{
1 + \frac{V_1(T_c)}{2a_0 G_2^{1/4}}} \,~  \Bigg)\, , 
\ee
with $a_0 = (3 \pi^2/768)^{1/4}$ in the case of three flavours.\\  
In their analysis, \citet{ST07} assume  $V_1(T_c) = 0.5 \,{\rm GeV}$, 
thus, $T_c$ in equation (\ref{Tc0}) is a simple function of $G_2$, and is represented 
in Fig.~\ref{Tc_G2} by the curve labelled $V_1(T_c) = 0.5 \,{\rm GeV}$.   
This result can hence be compared with lattice QCD calculations of $T_c$ giving 
the possibility of extracting the range of values for the gluon condensate compatible 
with lattice results. This comparison has been done by \citet{ST07}, 
and it is done in this work in Fig.~\ref{Tc_G2}, where we consider recent 
lattice QCD calculations of $T_c$ \citep{bors10,baz12}.  
Details to the specific lattice QCD calculations are given in the caption to Fig.~\ref{Tc_G2}.  
As one can see, the comparison with lattice QCD calculations of $T_c$ restricts 
the gluon condensate in a rather narrow range  $G_2 = 0.0025$--$0.0050 \, \rm{GeV}^4$. 

Next, to verify if these values of $G_2$ are compatible with those extracted in 
Section~\ref{NSstr} from hybrid star calculations and measured neutron star masses, 
we need to relate the parameter $V_1 \equiv V_1(0)$, entering in the zero temperature 
EOS of the quark phase, with  $V_1(T_c)$ in equation~(\ref{Tc0}). 
To this end, one can integrate equation~(\ref{v1}) using the non-perturbative 
contribution of equation~(\ref{d1nonpt}) to the colour electric correlator $D_1^E(x)$ and assuming 
that the normalization factor $D_1^E(0)$ does not depend on temperature. 
The latter assumption is supported, up to temperatures very near to $T_c$,  
by lattice calculations \citep{pisa}.   
Therefore, one gets
\be
\label{v1_T}
V_1(T)  = V_1(0) 
\bigg\{1-\frac{3}{2} \frac{\lambda T}{\hbar c} + 
\frac{1}{2} \bigg(1 + 3 \frac{\lambda T}{\hbar c}\bigg) e^{- \frac{\hbar c}{\lambda T}} 
\bigg\}  \, . 
\ee
Thus, $V_1(T_c) = 0.5 \,{\rm GeV}$ \citep{ST07} corresponds to  $V_1(0) = 0.85 \,{\rm GeV}$ 
to be used in the $T = 0$ EOS of the quark phase. 
In this case, as we found in Section~\ref{NSstr}, no phase transition occurs in neutron 
stars (i.e. $\rho_1 > \rho_c^{NS}$) for all the considered values of $G_2$. 
Thus, for these values of the EOS parameters PSR~J1614--2230 would be a pure nucleonic star. 
  
We can also evaluate the FCM transition temperature at $\mu_b=0$ corresponding 
to the case  $V_1(0) = 0.01 \,{\rm GeV}$ used in Section~\ref{NSstr} for hybrid star 
calculations with the $T=0$ FCM EOS.      
To this purpose, we solve numerically equations~(\ref{Tc0}) and (\ref{v1_T}), 
and we obtain the results represented in Fig.~\ref{Tc_G2} by the curve labelled 
$V_1 = 0.01 \,{\rm GeV}$. The comparison of these results with 
lattice QCD calculations \citep{bors10,baz12} of $T_c$ restricts the gluon 
condensate in the range  $G_2 = 0.0103$--$0.0180 \, \rm{GeV}^4$.  
Coming now to the astrophysical constraints on the gluon condensate,  
the vertical green line in Fig.~\ref{Tc_G2}  represents the lower limit for $G_2$ 
which is compatible, in the case $V_1(0) = 0.01 \,{\rm GeV}$, with the lower bound 
of the measured mass of PSR~J1614--2230 (see Section~\ref{NSstr}).\\ 
A similar analysis can be done for the case $V_1(0) = 0.10 \,{\rm GeV}$. 
Now the comparison between the FCM transition temperature at $\mu_b=0$ (curve 
labelled $V_1 = 0.10 \,{\rm GeV}$ in Fig.~\ref{Tc_G2}) and lattice QCD 
calculations of the same quantity gives 
$G_2 = 0.0085$--$0.0153 \, \rm{GeV}^4$, 
whereas one gets $G_2  \geq 0.006 \, {\rm GeV}^4$ from the lower bound of 
the measured mass of PSR~J1614--2230.

\section{Conclusions}

In this Letter, we have studied the hadron--quark deconfinement transition 
in $\beta$-stable nuclear matter and the structural properties of hybrid stars using an EOS 
for the quark phase derived from the FCM extended to finite baryon chemical potential.  
We obtained stable hybrid star configurations for all the values of the gluon condensate 
fulfilling the condition $\rho_1(G_2) < \rho_c^{NS}(G_2)$, i.e. the deconfinement 
transition can occur in pure nucleonic stars.   

We have established that the values of the gluon condensate extracted within the FCM 
from lattice QCD calculations of the deconfinement transition temperature at $\mu_b = 0$ 
are fully consistent with the value of the same quantity derived by the mass measurement 
of PSR~J1614--2230.  
The FCM thus provides a powerful tool to link numerical calculations of QCD on a 
space--time lattice with neutron stars physics.


\section*{Acknowledgements}

It is a pleasure to thank Claudio Bonati for very useful discussions.

\label{lastpage}

\end{document}